\documentstyle{article}
\begin{document}
\title{CONSISTENT CONSTRUCTION OF REALISTIC ONE-BODY DENSITY MATRIX IN
NUCLEI}
\author{A.N. Antonov$^a$, S.S. Dimitrova$^a$, M.K. Gaidarov$^a$,
        M.V. Stoitsov$^a$,\\
M.E. Grypeos$^b$, S.E. Massen$^b$, K.N. Ypsilantis$^b$}
\date{$^a${\it Institute of Nuclear Research and Nuclear Energy,\\
       Bulgarian Academy of Sciences, Sofia 1784, Bulgaria}\\
$^b${\it Department of Theoretical Physics, Aristotle University of\\
Thessaloniki, Thessaloniki 54006, Greece}}
\maketitle

\begin{abstract}
A phenomenological method based on the natural orbital representation is
applied to construct the ground state one-body density matrix which describes
correctly both density and momentum distributions in $^{4}He$, $^{16}O$ and
$^{40}Ca$ nuclei. The parameters of the matrix are fixed by a best fit to the
experimental density distribution and to the correlated nucleon momentum
distribution. The method allows the natural orbitals, the occupation
probabilities and the depletion of the Fermi sea to be obtained.
Ground-state characteristics of $^{4}He$, $^{16}O$ and $^{40}Ca$ nuclei, such
as rms radii and mean kinetic energies are calculated, as well.
\end{abstract}

\section{Introduction}
The one-body density matrix (OBDM) plays an important role in the nuclear
structure theory. It is known that the nuclear wave function contains much
more information than it is accessible to observation \cite{1,2}. All
measurable characteristics of the nuclear ground state (apart from
the energy) correspond to single-particle operators whose expectation
values can be expressed in terms of the OBDM elements. The ground state energy
can be calculated by a sum rule analysis \cite{3,4,5} also by means of
the OBDM. Thus the determination of the OBDM from first principles
\cite{6} is one of the most essential aims of the nuclear theory.

As shown in \cite{7,8} (and discussed in \cite{9,10,11,12}), an appropriate
criterion for the proximity of the OBDM $\rho_{0}(\bf r, \bf r ^{\prime})$
corresponding to a single-Slater determinant wave function (i.e.
$\rho_{0}^{2}$=$\rho_{0}$, for instance in the Hartree-Fock approximation) to
the OBDM $\rho(\bf r, \bf r ^{\prime})$ of the true (correlated) ground state
is that the "mean-square deviation per particle"
\begin{equation}
\sigma=A^{-1}Tr[(\rho - \rho_{0})^{2}]
\label{1}
\end{equation}
(where $A$ is the mass number) should be minimal. It has been shown in \cite{8}
that $\sigma $ is minimal when $\rho_{0}(\bf r, \bf r ^{\prime})$ corresponds
to
single Slater determinant wave function constructed with natural orbitals,
i.e. with the single-particle wave functions which diagonalize the OBDM
$\rho (\bf r, \bf r ^{\prime})$ \cite{13}. As was shown in \cite{9}, if in the
Hartree-Fock approximation the OBDM diagonal elements $\rho_{0}^{HF}(\bf r,
\bf r ^{\prime})$ are fitted to reproduce correctly the exact density
distribution $\rho (\bf r)\equiv \rho (\bf r, \bf r)$ (i.e. $\rho_{0}^{HF}
(\bf r,\bf r)\simeq \rho (\bf r)$), the extremum property
$\sigma $=$\sigma_{min}$ leads inevitably to an increasing deviation between
the non-diagonal elements of these two matrices ($\rho (\bf r, \bf r
^{\prime})$
and $\rho_{0}^{HF}(\bf r,\bf r ^{\prime})$ at $\bf r \neq \bf r ^{\prime} $).
Th
non-diagonal elements are related, however, to the nucleon momentum
distribution (NMD):
\begin{equation}
n(\bf k )=\int \rho ({\bf r},{\bf r^{\prime}})\exp[i{\bf k}({\bf r}-{\bf
r^{\prime}})]d\bf r d\bf r^{\prime} .
\label{2}
\end{equation}
This leads to the conclusion \cite{9} that generally the mean-field
approximation
is unable to give simultaneously a correct description of the two basic nuclear
ground state characteristics, namely the density and momentum distributions.
This has been supported by the calculations in the case of $^{4}He$ nucleus
\cite{14}.

The role of the nucleon-nucleon correlations in the nuclear systems can be seen
by the analyses of the quantity $\sigma $ \cite{10}. While in the Hartree-Fock
approximation $\sigma $ vanishes and the contribution of the random-phase
approximation to $\sigma $ is about $0.002$, for realistic N-N forces the
short- and medium-range correlation effects lead to a value of
$\sigma_{min}\simeq 0.02\div 0.03$ that implies a Fermi sea depletion of about
$10\div 15\%$. The latter is confirmed by the experimental data \cite{15,16}.
The short-range and tensor correlations are responsible for the
existence of high-momentum components in the realistic momentum distributions
in nuclei which is not the case in the mean-field approximation.

A phenomenological method to construct a more realistic OBDM is suggested in
\cite{1,2} for the cases of $^{16}O$ and $^{40}Ca$ nuclei, both with fractional
occupation numbers and full occupation of the first $A$ natural orbitals. The
latter are expanded in terms of harmonic-oscillator wave functions. The
expansion
coefficients, as well as the occupation numbers are fixed by a best fit to the
experimental density distributions of $^{16}O$ and $^{40}Ca$ minimizing the rms
percentage deviation between the experimental and trial radial density. A
satisfactory description of g.s. properties, such as the g.s. energy and rms
radii, is obtained using the OBDM constructed within this method, in
contrast with the predictions of the Hartree-Fock approximation.

Here we would like to emphasize that according to the discussion given above,
the OBDM obtained in the method from \cite{1,2} cannot be a realistic one
because
it is obtained by a best fit only of its diagonal elements to the experimental
density distribution (which is not very sensitive to the N-N correlations) and
is not aimed to describe the nucleon momentum distribution which is related
also to the non-diagonal elements of the OBDM and is mostly affected by the
presence of short-range and tensor N-N correlations in the nuclear system.

The aim of this work is to construct the g.s. OBDM in $^{4}He$, $^{16}O$ and
$^{40}Ca$ nuclei in a consistent way following the method from
\cite{1,2} but providing in an optimal way a correct
description of both density and momentum distributions in nuclei considered.
Solving this problem, we show that, in principle, the expansion of the natural
orbitals in terms of harmonic oscillator functions (in the truncated s.p.
space) cannot give a correct description of the realistic high-momentum
components of the nucleon momentum distribution. This imposes to use in
the expansion of the natural orbitals another set of s.p. functions, for
instance that one corresponding to infinite square-well potential.

The method suggested in this work enables us to construct a more realistic OBDM
which includes that part of the nucleon correlations which are responsible for
the high-momentum components of the nucleon momentum distribution and is
consistent with the nuclear density distribution.

Some basic relations are given in Section 2. The phenomenological
expression for the OBDM is presented in Section 3. The results of the numerical
calculations are given and discussed in Section 4.

\section{Nuclear ground-state properties}

In order to evaluate the single-particle characteristics of the nuclear ground
state one needs the OBDM which is defined by the expression:
\begin{equation}
\rho ({\bf r},{\bf r^{\prime} })=A \int \Psi ^{*}( {\bf r},{\bf r}_{2},...)
\Psi ({\bf r^{\prime} },{\bf r}_2,...) d{\bf r}_{2}d{\bf r}_{3}...d{\bf
r}_{A},
\label{3}
\end{equation}
where $\Psi ({\bf r}_{1},{\bf r}_{2},...,{\bf r}_{A})$ is the normalized
$A$- nucleon ground-state wave function. The integration in Eq.(3) is carried
out over the radius vectors and summation over spin and isospin variables
is implied.

The one-body density matrix has the simplest form in the so called natural
orbital representation \cite{13}:
\begin{equation}
\rho ({\bf r},{\bf r^{\prime}})=\sum_{i} \lambda_{i}\psi_{i}^{*}({\bf
r})\psi_{i} ({\bf r^{\prime} }),
\label{4}
\end{equation}
where $\lambda_{i}$ are the occupation numbers (eigenvalues of $\rho$)
corresponding
to the natural orbitals $\psi_{i}({\bf r})$ (eigenfunctions of $\rho$) which
form a complete orthonormal set. The values of $\lambda_{i}$ satisfy the
general
conditions: $0\leq \lambda_{i} \leq 1$ and $\displaystyle \sum_{i}
\lambda_{i}=A$.
Usually there are $A$ orbitals $\psi_{i}({\bf r})$ for which the occupation
probabilities $\lambda_{i}$ are significantly larger than those for the
others. As
in the mean field approach, these are called the hole-state orbitals while the
others are called particle-state orbitals \cite{17}.

In the momentum space the OBDM (Eq.(3)) reads
\begin{equation}
n({\bf k},{\bf k^{\prime}})\equiv A\int \tilde{\Psi}^{*}({\bf k},{\bf k}_{2},
...,{\bf k}_{A})\tilde{\Psi}({\bf k}^{\prime},{\bf k}_{2},...,{\bf k}_{A})
d{\bf k}_{2}...d{\bf k}_{A},
\label{5}
\end{equation}
where $\tilde{\Psi}({\bf k}_{1},{\bf k}_{2},...,{\bf k}_{A})$ is the Fourier
transform of the $A$-nucleon wave function $\Psi({\bf r}_{1},{\bf
r}_{2},...,{\bf r}_{A})$. Its diagonal elements determine the nucleon momentum
distribution (Eq.(2)). The associated natural orbital representation
\begin{equation}
n({\bf k},{\bf k^{\prime} })=\sum_{i} \lambda_{i}\tilde{\psi}_{i}^{*}({\bf k})
\tilde{\psi}_{i}({\bf k^{\prime}})
\label{6}
\end{equation}
depends on the same occupation numbers $\lambda_{i}$ entering Eq.(4) and the
Fourier
transform of the natural orbitals $\psi_{i}({\bf r})$ in the momentum space
\begin{equation}
\tilde{\psi}_{i}({\bf k})=(2\pi)^{-3/2}\int \psi_{i}({\bf r})\exp(i{\bf k}
{\bf r})d{\bf r}.
\label{7}
\end{equation}
For the purposes of our work a particular attention will be paid to the local
density distribution (the diagonal elements of the OBDM in the coordinate
space):
\begin{equation}
\rho({\bf r})\equiv \rho({\bf r},{\bf r})=\sum_{i} \lambda_{i}|\psi_{i}({\bf
r})|^{2} \label{8}
\end{equation}
and the nucleon momentum distribution (the diagonal elements of the OBDM in the
momentum space):
\begin{equation}
n({\bf k})\equiv n({\bf k},{\bf k})=\sum_{i} \lambda_{i}|\tilde{\psi}_{i}({\bf
k})|^{2}.
\label{9}
\end{equation}
The second moments of $\rho ({\bf r})$ and $n({\bf k})$ define respectively the
rms radius of the nucleus
\begin{equation}
\langle r^{2} \rangle ^{1/2}=\left \{A^{-1}\int \rho ({\bf r})r^{2}d{\bf
r}\right \}^{1/2}
\label{10}
\end{equation}
and its mean kinetic energy
\begin{equation}
\langle T \rangle =\frac{\hbar ^{2}}{2m}\int n({\bf k})k^{2}d{\bf k}.
\label{11}
\end{equation}

The knowledge of a realistic one-body density matrix would allow to describe
correctly the nuclear characteristics, such as the nucleon momentum and density
distributions, rms radii and mean kinetic energies taking into account the
effects of the nucleon-nucleon correlations in the nuclear system.

\section{Phenomenological method for the constructing of the one-body density
matrix}

In this Section a phenomenological expression for the OBDM is given. The
parameters in this expression can be obtained by a numerical procedure by
fitting
the local density $\rho (r)$ and the momentum distribution $n(k)$ to the
corresponding experimental data (or to realistic theoretical estimations)
\cite{18,19,20}.

For closed-shell nuclei with equal numbers of protons and neutrons it is
reasonable to assume that the protons and neutrons have the same one-body
density matrix. For such nuclei (with total spin $J=0$) the OBDM has to be
diagonalized in the $\{ljm\}$ subspace of the complete space of the natural
orbitals \cite{21} ($l$,$j$,$m$ being the quantum numbers corresponding to
the s.p. orbital and total momentum and its projection). Due to the spherical
symmetry of the considered nuclei the natural orbitals can be looked for in the
form:
\begin{equation}
\psi_{i}({\bf r})\equiv \psi_{nlm}({\bf r})=R_{nl}(r)Y_{lm}(\theta,\varphi ).
\label{12}
\end{equation}
It is known that the radial part of the natural orbitals $R_{nl}(r)$ and
especially of the particle-state orbitals differ from the single-particle wave
functions obtained within the mean field approximation to the ground state of a
fermion system \cite{17,18,19,22}. Following exactly the method suggested in
\cite{1,2}, we expand the radial part of the natural orbitals of the OBDM in
terms of three single-particle functions preserving all the usual symmetries
for spherical nuclei:
\begin{equation}
R_{a}(r)=\sum_{i=1}^{3} C_{i}^{a}\varphi_{il}(r),\;\;\;\; (a\equiv nl)
\label{13}
\end{equation}
In Eq.(13) $\{\varphi_{nl}(r)\}$ is a set of orthonormal s.p. wave functions.
The expansion coefficients $C_{i}^{a}$ are fitting parameters satisfying the
orthonormalization conditions. For the numerical calculations it is convenient
to re-express them using polar coordinates:
\begin{eqnarray}
C_{1}^{a} &=& \cos \theta_{a} \nonumber \\
C_{2}^{a} &=& \sin \theta_{a} \cos\varphi_{a}\\
C_{3}^{a} &=& \sin \theta_{a} \sin\varphi_{a}.\nonumber
\label{14}
\end{eqnarray}
In this way the normalization is automatically guaranteed. The orthogonality
among states of the same symmetry reduces the number of parameters involved in
Eq.(14).

By means of Eqs.(12-14) one can obtain the expression for the one-body density
matrix (Eq.(4)) within this method. The diagonal elements of the OBDM in
coordinate space determine the local density distribution:
\begin{equation}
\rho (r)=\frac{1}{4\pi} \sum_{nl} 4(2l+1)\lambda_{nl}|R_{nl}(r)|^{2}.
\label{15}
\end{equation}
The occupation numbers $\lambda_{nl}$ satisfy the condition:
\begin{equation}
\sum_{nl} 4(2l+1)\lambda_{nl}=A
\label{16}
\end{equation}
and the normalization of $\rho (r)$ is:
\begin{equation}
\int \rho ({\bf r})d{\bf r}=A.
\label{17}
\end{equation}
An expression similar to that for $\rho (r)$ determines the nucleon momentum
distribution:
\begin{equation}
n(k)=\frac{1}{4\pi}\sum_{nl} 4(2l+1)\lambda_{nl}|\tilde{R}_{nl}(k)|^{2};
\;\;\;\; \int n({\bf k})d{\bf k}=A.
\label{18}
\end{equation}
In Eq.(18)
\begin{equation}
\tilde{R}_{a}(k)=\sum_{i=1}^{3} C_{i}^{a}\tilde{\varphi}_{il}(k),
\;\;\;\; (a\equiv nl)
\label{19}
\end{equation}
where $\tilde{\varphi}_{nl}(k)$ are the orthonormalized s.p. wave functions in
the momentum space.
Obviously, due to the truncation of the expansion (13) (and the corresponding
one (19)) up to $i$=3, the radial part of the natural orbitals depends strongly
on the choice of the single-particle wave functions $\{\varphi_{nl}\}$. We use
three sets of s.p. wave functions $\{\varphi_{nl}\}$ corresponding to: 1) the
harmonic-oscillator potential (HO):
\begin{equation}
V(r)=-V_{0}+\frac{1}{2}mw^{2}r^{2} \;\;\;\; (V_{0}>0),
\label{20}
\end{equation}
2) the square-well potential with infinite walls (SW):
\begin{equation}
 V(r) = \left\{ \begin{array}{ll}
-V_{0},        & \mbox{$r<x,\;\;\;V_{0}>0$}\\
\;\;\; \infty, & \mbox{$r>x$}
\end{array}
\right.
\label{21}
\end{equation}
and 3) the modified harmonic-oscillator potential (MHO) \cite{23,24,25}:
\begin{equation}
V(r)=-V_{0}+\frac{1}{2}mw^{2}r^{2}+\frac{B}{r^{2}},\;\;\;V_{0}>0,B\geq 0.
\label{22}
\end{equation}
The modified harmonic-oscillator potential (Eq.(22)) behaves like a
harmonic-oscillator one for large values of $r$ but it has in addition a
repulsive term ($B$ determines the strength of the repulsion) which is the
dominant one at short distances from the origin. One can expect that this term
simulates to some extent the effects of strong repulsion in the nucleon-nucleon
interaction at small distances. The single-particle potential (22) had been
suggested and used \cite{23,24,25} for the study of charge formfactors and
nucleon momentum distributions of light nuclei. One of the main advantages of
this potential is that analytical expressions are derived for the
single-particle wave functions and for other useful quantities (see e.g.,
\cite{25}). Though the MHO potential (22) leads to an infinite repulsion at the
origin ($r=0$), the corresponding single-particle wave functions which are used
in the present work are correct for all values of $r$.

In the natural orbital representation the local density and the momentum
distributions
(Eqs.(15) and (18)) are expressed as sums over the whole single-particle space.
In the numerical  calculations we include in these sums all hole-state orbitals
and several particle-state orbitals. In this way the expression for the OBDM is
determined and it can be used within a minimization procedure which leads to
the constructing of a realistic one-body density matrix.

\section{Numerical calculations. Results and discussion}

The phenomenological method for constructing the realistic one-body density
matrix given in the previous Section is applied to the case of the
double-closed shell nuclei $^{4}He$, $^{16}O$ and $^{40}Ca$. Obviously, the
expansion (4)
should be truncated. In the case of $^{4}He$ and $^{16}O$ we include the $1s$,
$1p$, $1d$
and $2s$ states and for $^{40}Ca$ nucleus-the $1s$, $1p$, $1d$, $2s$, $2p$,
$1f$ and $1g$ states. The parameters of the OBDM, namely the occupation numbers
$\{\lambda_{nl}\}$, the coefficients $\{C_{i}^{nl}\}$ and the parameters of the
sets of the orthonormal single-particle functions corresponding to the
potentials (20)-(22) (the oscillator parameter $\alpha=(mw/\hbar)^{1/2}$,
the radius of the well $x$ and the parameter $b=8mB/\hbar^{2}$) are fitting
parameters. The values of the parameters are fixed by a best fit of the
diagonal elements of the OBDM in the coordinate space and in the momentum
space (Eqs.(15) and (18)) to the experimental local density $\rho (r)$
obtained in \cite{26,27,28,29} and to the nucleon momentum distribution $n(k)$.
In the case of $^{4}He$ we use the experimental data for $n(k)$ obtained by
means of the $y$-scaling analysis \cite{20} relying on the assumption that the
$1/q$ expansion is valid. Theoretical estimations for $n(k)$ in $^{16}O$ and
$^{40}Ca$ obtained within the Jastrow correlation method (JCM) \cite{18,19,30}
are used instead of experimental ones due to the lack of empirical data for
these nuclei. Though the low-order approximation to the JCM has some
deficiences, we use it due to the obtained simple analytical expressions for
the one-body nuclear characteristics, in particular, for the $n(k)$. The
latter reproduce the results from the complicated exp(S)-calculations for
$^{16}O$ \cite{31} including the high-momentum tail of $n(k)$. It is possible
also more realistic theoretical predictions for $n(k)$ in the nuclei
considered to be used.

Our fitting criterion is to minimize the relative deviation between the
experimental and theoretical local density distributions
\begin{equation}
\chi_{D}^{2}=\sum_{q} \left [ \frac{\rho_{q}^{exp}(r)-\rho_{q}^{th}(r)}
{\rho_{q}^{exp}(r)}\right ] ^{2},
\label{23}
\end{equation}
where $q$ is the number of points under which minimization is performed, and
between the theoretical nucleon momentum distribution for $^{16}O$ and
$^{40}Ca$ (or experimental $n(k)$ for $^{4}He$) and that corresponding to the
OBDM proposed in this work. As far as the latter is concerned it is preferable
to consider $\lg n(k)$ rather than $n(k)$ because the values of $n(k)$ rapidly
decrease by several orders of magnitude with the increase of the momentum $k$:
\begin{equation}
\chi_{M}^{2}=\sum_{q} \left [ \frac{(\lg n^{JCM}(k))_{q}-(\lg n^{th}(k))_{q}}
{(\lg n^{JCM}(k))_{q}} \right]^{2}.
\label{24}
\end{equation}
Otherwise, our minimization criterion would not be sensitive to both parts
(low-momentum and high-momentum region) of the nucleon momentum distribution.
Taking into account Eqs.(23) and (24) we minimize the expression:
\begin{equation}
F=\sqrt{\chi _{D}^{2}} + \sqrt{\chi _{M}^{2}}.
\label{25}
\end{equation}
The minimization is accomplished using the program MINUIT. There have
been involved in it 12 (or 13 when MHO s.p. wave functions are used) fitting
parameters in the case of $^{4}He$ and $^{16}O$ and 20 parameters for
$^{40}Ca$ nucleus.

The quality of the minimization procedure is illustrated in Table 1. There are
listed the values of $\chi_{D}^{2}$ (Eq.(23)), $\chi_{M}^{2}$ (Eq.(24)) per
fitting point $q$ and $F$ (Eq.(25)). For all nuclei examined it is
seen that the minimal value of $F$ is achieved using square-well s.p. wave
functions. For example, in the case of $^{16}O$ the local density distribution
(15) and the nucleon momentum
distribution (18) are approximated with almost the same accuracy. A low value
of $F$ is also obtained when modified harmonic-oscillator s.p. wave functions
are used but $\chi_{D}^{2}/q$ and $\chi_{M}^{2}/q$ differ by two orders of
magnitude. It means that in this case the local density distribution is
described better then
the nucleon momentum distribution. As is well known, the
harmonic-oscillator s.p. wave functions are appropriate to describe the local
density distribution of $^{16}O$ within the simple shell model. Considering the
values of $\chi_{D}^{2}$, $\chi_{M}^{2}$ and $F$ one can see
that the proposed method in this case is unable to reproduce
simultaneously the local density and the
nucleon momentum distribution. Similar calculations can be drawn considering
the minimization procedure for $^{40}Ca$. The best fit value of $F$ is obtained
using
square-well s.p. wave functions. The local density distribution and the nucleon
momentum distribution are described satisfactorily to almost the same accuracy.

The values of the occupation numbers $\lambda_{nl}$, of the coefficients
$C_{i}^{nl}$ and of the parameters of the s.p. wave functions in the expansion
of the OBDM for $^{4}He$, $^{16}O$ and $^{40}Ca$ are given in Tables 2, 3 and
4, respectively. Thus the one-body density matrix is fully determined within
the method proposed. The values of the occupation numbers $\lambda_{nl}$ are
compared with the occupation numbers $\eta_{nl}$ \cite{19} obtained within the
natural orbital representation of the OBDM in the low-order approximation of
the Jastrow correlation method which can be considered as an "exact limit" of
the method proposed in the present paper. It is seen that the values of
$\lambda_{nl}$ in the case of square-well s.p. wave functions are in good
agreement with the natural occupation numbers $\eta_{nl}$.

The depletion of the Fermi sea in $^{4}He$ is 18.6\%, 13.2\% and 11.3\%
in the cases when harmonic-oscillator, modified harmonic-oscillator-  and
square-well s.p. wave functions are used, respectively, in the expansions
(13) and (19).
Its corresponding values using the same types of s.p. wave functions in the
case of $^{16}O$ are 18\%, 16.6\% and 12.4\%, respectively. In the case of
$^{40}Ca$ the depletion is 20\% and 15\% when harmonic-oscillator and
square-well s.p. wave functions are used, respectively. It can be seen again
that the best description of the OBDM within the proposed procedure is achieved
when square-well s.p. wave functions are used because the depletion in this
case is minimal and it is the closest to the corresponding value in the natural
orbital representation \cite{19}--6\% for $^{4}He$, 4\% for $^{16}O$ and
6.3\% for $^{40}Ca$ and to the experimental one \cite{32}--9.4\% for $^{40}Ca$.
We would like to mention here that the differences of our results from those
obtained in \cite{19} and from the experimental ones for $^{40}Ca$ \cite{32}
have to be considered bearing in mind the deficiences of the low-order
approximation to the Jastrow correlation method used in \cite{19}, the model
dependence of the analyses of the empirical data and, of course, the
limitations of the suggested method. Nevertheless, we emphasize that our
results are within the theoretical limits for the depletion of the Fermi
sea which is expected to be approximately between 5 and 15\%.

In Figs.1 and 2 the results for the local density and momentum distribution in
$^{4}He$, $^{16}O$ and $^{40}Ca$ obtained by means of the one-body density
matrix
determined in the model are presented. The best solution is reached by using of
square-well s.p. wave functions. As is seen from Fig.1 an acceptable
quantitative agreement with the experimental data for the density distribution
$\rho (r)$ (at least in the surface region) can be achieved almost
independently on the choice of the s.p. wave function sets. However, only
the use of the SW s.p. wave functions corresponding to the potential (21) gives
a correct description of the realistic high-momentum components of the nucleon
momentum distribution (see Fig.2). As is expected, the use of HO s.p. wave
functions does not reproduce realistically $n(k)$. It can be also seen
from Figs.2b and 2c that although the OBDM obtained in the method from
\cite{1,2} gives a correct description of the density distribution,
it is unable to describe the high-momentum behaviour of the momentum
distribution which is sensitive to the short-range N-N correlations.
In the cases of $^{4}He$ and $^{16}O$ (Figs.2a and 2b) the use of MHO s.p. wave
functions corresponding to the potential (22) leads to an improvement of the
behaviour of $n(k)$ at higher momenta which is due to the existence of the
repulsive term in the potential.

In Table 5 the calculated rms radii $\langle r^{2} \rangle^{1/2}$ (Eq.(10)) and
mean kinetic energies per nucleon $\langle T \rangle /A$ (Eq.(11)) of $^{4}He$,
$^{16}O$ and $^{40}Ca$ and their values deduced from the JCM \cite{30} are
presented. The values of the rms radii for all nuclei examined are in accord
with the corresponding experimental values \cite{26,33}. Due to the fact
mentioned above that the expansion of the natural orbitals in terms of HO
functions cannot give a correct description of the realistic high-momentum
components of the nucleon momentum distribution (see Fig.2c), it can be seen
in Table 5 a quite increased value of $\langle T \rangle /A$ for $^{40}Ca$
in comparison with the result obtained by using SW s.p. wave function set.
Therefore, we have demonstrated that the SW s.p. wave functions are the
best ones among the other used in the sense of correct description of both
density and momentum distributions in $^{4}He$, $^{16}O$ and $^{40}Ca$. Here
we would like to mention that our choice of the truncation in the expansions
(13) and (19) up to $i=3$ was imposed from one side by the desire to follow
exactly the method previously suggested in \cite{1,2} in order to compare our
results with those from the mentioned works. On the other hand, it is
well-known that the expansions (13) and (19) in terms of the complete set
of s.p. wave functions would be exact expressions for the natural
orbitals. Of course, this is impossible to be done and in the practical
applications we have to find a proper value of $i$ which would lead to  a
reasonable number
of the minimization parameters. The inclusion of higher terms in the expansions
($i>3$) would make the results of the method less sensitive to the particular
choice of the s.p. wave function set, but at the same time this would increase
enormously the difficulties of the practical minimization procedure. Hence,
the applications of the method would be hardly possible. In our opinion, the
choice $i=3$ is an acceptable compromise within the method.

\section{Conclusions}

The results of this work can be summarized as follows:\\
i) a phenomenological
method for a consistent construction of the one-body density matrix is
suggested. In contrast to the method from \cite{1,2}, our procedure allows to
obtain an one-body density matrix which gives realistically both the local
density and the nucleon momentum distribution of $^{4}He$, $^{16}O$ and
$^{40}Ca$ nuclei.\\
ii) The method gives the natural orbitals, the occupation probabilities and the
depletion of the Fermi sea in the nuclei considered.\\
iii) The rms radii and mean kinetic energies of $^{4}He$, $^{16}O$ and
$^{40}Ca$ are calculated with the method.\\
iv) Consideration of both non-diagonal and diagonal elements of $\rho ({\bf
r},{\bf r^{\prime}})$ makes it possible to treat in a
consistent way the effects of the N-N short-range correlations in nuclei on the
one-body density matrix and the related ground state characteristics.
\vspace{1.3cm}\\
{\Large\bf Acknowledgements}
\vspace{0.4cm}\\
The authors are deeply grateful to Professor S. Boffi for the valuable
discussions and to Professor C. Ciofi degli Atti for providing the
experimental data for $n(k)$ in $^{4}He$ from $y$-scaling analysis. One of us
(S.S.D.) would like to thank Prof. M. Grypeos for the kind hospitality at the
Aristotle University of Thessaloniki. This work was partly supported by the
EEC under Contract ERB-CIPA-CT-92-2231 and by the Bulgarian National
Science Foundation under the Contract Nr.$\Phi$--406.

\vspace{2cm}
%
\noindent
{\Large\bf Figure Captions}\vspace{.4cm}\\
Figure 1. Local density distribution in $^{4}He$ (a), $^{16}O$ (b) and
$^{40}Ca$ (c) obtained by using different sets of s.p. wave functions in
Eq.(15). The normalization is
$\displaystyle \int \rho ({\bf r})d{\bf r}=A/2$.
\vspace{1cm}\\
Figure 2. Nucleon momentum distribution in $^{4}He$ (a), $^{16}O$ (b) and
$^{40}Ca$ (c) obtained by using different sets of s.p. wave functions in
Eq.(18). The normalization is $\displaystyle \int n({\bf k})d{\bf k}=1$.

\newpage
\noindent
Table 1. $\chi^{2}$-values per fitting point for the local density (Eq.(23))
and the nucleon momentum distribution (Eq.(24)) and the $F$-value (Eq.(25)) for
$^{4}He$, $^{16}O$ and $^{40}Ca$.
\begin{center}
\begin{tabular}{ccccccc} \hline \hline
Nuclei & S.p. wave & $\chi_{D}^{2}/q$ & $\chi_{M}^{2}/q$ & $F$\\
& functions & & & \\
\hline
$^{4}He$   & HO          &  0.0040  &  0.0379  &  1.204 \\
           & MHO         &  0.0031  &  0.0260  &  1.027 \\
           & SW          &  0.0068  &  0.0036  &  0.932 \\
\hline
$^{16}O$   & HO          &  0.0098  &  1.2458  &  6.239 \\
           & MHO         &  0.0019  &  0.1271  &  2.104 \\
           & SW          &  0.0199  &  0.0170  &  1.888 \\
\hline
$^{40}Ca$  & HO          &  0.1805  &  0.6678  &  8.147 \\
           & SW          &  0.0168  &  0.0352  &  2.190 \\
\hline \hline
\end{tabular}
\end{center}

\newpage
\noindent
Table 2. Occupation numbers $\lambda_{nl}$, expansion coefficients $C_{i}^{nl}$
($i$=1,2,3), parameters of the s.p. wave functions and natural occupation
numbers $\eta_{nl}$ \cite{19} for $^{4}He$.
\begin{center}
\begin{tabular}{cccccccccc} \hline \hline
& \multicolumn{2}{c}{HO s.p. wave functions} & & \multicolumn{2}{c}{MHO s.p.
wave functions} & & \multicolumn{2}{c}{SW s.p. wave functions} \\
\cline{2-3}  \cline{5-6} \cline{8-9}
$nl$ & \multicolumn{2}{c}{$\alpha =0.648$ fm$^{-1}$} & &
\multicolumn{2}{c}{$\alpha =0.687$ fm$^{-1} \;\;b=0.511$} & &
\multicolumn{2}{c}{$x=3.70$ fm} & $\eta_{nl}$ \\
\cline{2-3}  \cline{5-6} \cline{8-9}
& $\lambda _{nl} $ & $C_{i}^{nl}$ & &$\lambda _{nl} $ & $C_{i}^{nl}$ & &
$\lambda _{nl} $ & $C_{i}^{nl}$ \\
\cline{1-3}  \cline{5-6} \cline{8-10}
     &       &\ 0.9806& &       &\ 0.9699& &       &\ 0.9755&\\
$1s$ &0.8138 &\ 0.1937& & 0.8679&\ 0.2435& &0.8873 &-0.2191 &0.9428\\
     &       & -0.0308& &       &\ 0.0001& &       &\ 0.0182&\\
\vspace{-.4cm}\\
     &       &-0.5893 & &       &-0.9643 & &       &\ 0.1411&\\
$1p$ & 0.0392&-0.5011 & &0.0145 &-0.1136 & &0.0063 &-0.9576 &0.0097\\
     &       &-0.6337 & &       &\ 0.2392& &       &\ 0.2513&\\
\vspace{-.4cm}\\
     &       &\ 0.1848& &       &-0.3959 & &       &-0.3405 &\\
$1d$ & 0.0033&\ 0.1256& &0.0064 &\ 0.1115& &0.0187 &\ 0.5185&0.0027\\
     &       &-0.9747 & &       &\ 0.9115& &       &-0.7843 &\\
\vspace{-.4cm}\\
     &       &\ 0.1319& &       &-0.1974 & &       &-0.1848 &\\
$2s$ & 0.0521&-0.7674 & &0.0566 &\ 0.7864& &0.0002 &-0.7222 &0.0027\\
     &       &-0.6275 & &       &\ 0.5854& &       &\ 0.6079&\\
\hline \hline
\end{tabular}
\end{center}

\newpage
\noindent
Table 3. Occupation numbers $\lambda_{nl}$, expansion coefficients $C_{i}^{nl}$
($i$=1,2,3), parameters of the s.p. wave functions and natural occupation
numbers $\eta_{nl}$ \cite{19} for $^{16}O$.
\begin{center}
\begin{tabular}{cccccccccc} \hline \hline
& \multicolumn{2}{c}{HO s.p. wave functions} & & \multicolumn{2}{c}{MHO s.p.
wave functions} & & \multicolumn{2}{c}{SW s.p. wave functions} \\
\cline{2-3}  \cline{5-6} \cline{8-9}
$nl$ & \multicolumn{2}{c}{$\alpha =0.604$ fm$^{-1}$} & &
\multicolumn{2}{c}{$\alpha =0.582$ fm$^{-1} \;\;b=0.731$} & &
\multicolumn{2}{c}{$x=4.60$ fm} & $\eta_{nl}$ \\
\cline{2-3}  \cline{5-6} \cline{8-9}
& $\lambda _{nl} $ & $C_{i}^{nl}$ & &$\lambda _{nl} $ & $C_{i}^{nl}$ & &
$\lambda _{nl} $ & $C_{i}^{nl}$ \\
\cline{1-3}  \cline{5-6} \cline{8-10}
     &       &\ 0.9174& &       &\ 0.9576& &       &\ 0.9967&\\
$1s$ &0.8002 &-0.2686 & & 0.9359&-0.0740 & &0.9940 &-0.0087 &0.9495\\
     &       &\ 0.2936& &       &\ 0.2783& &       &-0.0808 &\\
\vspace{-.4cm}\\
     &       &\ 0.9959& &       &-0.9928 & &       &\ 0.9919&\\
$1p$ & 0.8278&\ 0.0008& &0.8000 &-0.0035 & &0.8370 &-0.1251 &0.9646\\
     &       &-0.0899 & &       &\ 0.1194& &       &\ 0.0214&\\
\vspace{-.4cm}\\
     &       &\ 0.2225& &       &\ 0.5289& &       &\ 0.4743&\\
$1d$ & 0.1249&\ 0.8353& &0.0973 &\ 0.7562& &0.0655 &\ 0.1129&0.0057\\
     &       &\ 0.5028& &       &\ 0.3651& &       &\ 0.8731&\\
\vspace{-.4cm}\\
     &       &\ 0.1464& &       &\ 0.0780& &       &-0.0315 &\\
$2s$ & 0.0920&-0.4582 & &0.1774 &\ 0.9970& &0.1672 &\ 0.8750&0.0057\\
     &       &-0.8767 & &       &-0.0032 & &       &-0.4831 &\\
\hline \hline
\end{tabular}
\end{center}

\newpage
\noindent
Table 4. Occupation numbers $\lambda_{nl}$, expansion coefficients $C_{i}^{nl}$
($i$=1,2,3), parameters of the s.p. wave functions and natural occupation
numbers $\eta_{nl}$ \cite{19} for $^{40}Ca$.
\begin{center}
\begin{tabular}{ccccccc} \hline \hline
& \multicolumn{2}{c}{HO s.p. wave functions} & & \multicolumn{2}{c}{SW s.p.
wave functions}\\
\cline{2-3}  \cline{5-6}
$nl$ & \multicolumn{2}{c}{$\alpha =0.627$ fm$^{-1}$} & &
\multicolumn{2}{c}{$x=6.40$ fm} & $\eta_{nl}$\\
\cline{2-3} \cline{5-6}
& $\lambda _{nl} $ & $C_{i}^{nl}$ & & $\lambda _{nl} $ & $C_{i}^{nl}$\\
\cline{1-3}  \cline{5-7}
       &         &\ 0.5323 &&         & -0.5092&\\
$1s$   & 0.8027  &\ 0.7972 && 0.9227  &\ 0.8385&0.8898\\
       &         &-0.2849  &&         & -0.1939&\\
\vspace{-.4cm}\\
     &           &\ 0.9733 &&         &\ 0.9959&\\
$1p$ & 0.8000    & -0.0609 && 0.8718  & -0.0717&0.9376\\
     &           &-0.2214  &&         & -0.0560&\\
\vspace{-.4cm}\\
     &           &\ 0.8023 &&         &\ 0.8548&\\
$1d$ & 0.8005    & -0.5923 && 0.8013  & -0.5181&0.9463\\
     &           &\ 0.0746 &&         &\ 0.0335&\\
\vspace{-.4cm}\\
     &           & -0.8440 &&         &\ 0.8517&\\
$2s$ & 0.8020    &\ 0.5261 && 0.9558  &\ 0.4586&0.9579\\
     &           & -0.1045 &&         & -0.2536&\\
\vspace{-.4cm}\\
     &           &\ 0.2197 &&         & -0.0203&\\
$2p$ & 0.1005    &\ 0.5283 && 0.1686  & -0.7750&0.0168\\
     &           &\ 0.8201 &&         &\ 0.6317&\\
\vspace{-.4cm}\\
     &           &\ 0.9913 &&         &\ 0.2266&\\
$1f$ & 0.000002  & -0.1106 && 0.0465  &\ 0.2341&0.0127\\
     &           & -0.0717 &&         &\ 0.9454&\\
\vspace{-.4cm}\\
     &           &\ 0.0453 &&         &\ 0.0589&\\
$1g$ & 0.1879    &\ 0.5765 && 0.0742  & -0.2858&0.0087\\
     &           &\ 0.8158 &&         &\ 0.9565&\\
\hline \hline
\end{tabular}
\end{center}

\newpage
\noindent
Table 5. Rms radii and mean kinetic energies per nucleon of $^{4}He$, $^{16}O$
and $^{40}Ca$ calculated by using different s.p. wave function sets for the
construction of the one-body density matrix.
\begin{center}
\begin{tabular}{cccc} \hline \hline
Nuclei &  & $\langle r^{2} \rangle ^{1/2}$ &
$\langle T \rangle $/$A$\\
&  &  (fm)  &  (MeV) \\
\hline
$^{4}He$   & HO          &  1.66  &  12.33\\
           & MHO         &  1.65  &  12.93\\
           & SW          &  1.67  &  21.47\\
           & EXP [26]    &  1.67  &       \\
           & JCM [30]    &  1.60  &  25.86\\
\hline
$^{16}O$   & HO          &  2.70  &  20.08\\
           & MHO         &  2.71  &  16.29\\
           & SW          &  2.60  &  26.52\\
           & EXP [33]    &  2.71  &       \\
           & JCM [30]    &  2.63  &  21.34\\
\hline
$^{40}Ca$  & HO          &  3.46  &  38.29\\
           & SW          &  3.42  &  26.94\\
           & EXP [33]    &  3.48  &       \\
           & JCM [30]    &  3.50  &  23.36\\
\hline \hline
\end{tabular}
\end{center}

\newpage
\begin{thebibliography}{99}
\bibitem{1} S. Boffi and F.D. Pacati, Nucl. Phys. {\bf A204} (1973) 485
\bibitem{2} S. Boffi and F.D. Pacati, Instituto Lombardo (Rend.Sc.) {\bf A107}
(1973) 321
\bibitem{3} V.M. Galitskii and A.B. Migdal, JETP (Sov. Phys.) {\bf 7} (1958) 96
\bibitem{4} S. Boffi, Nuovo Cim. Lett. {\bf 1} (1971) 931
\bibitem{5} D.S. Koltun, Phys. Rev. Lett. {\bf 28} (1972) 182
\bibitem{6} A.J. Coleman, Rev. Mod. Phys. {\bf 35} (1963) 668
\bibitem{7} W. Kutzelnigg and V.H. Smith, J. Chem. Phys. {\bf 41} (1964) 896
\bibitem{8} D.H. Kobe, J. Chem. Phys. {\bf 50} (1969) 5183
\bibitem{9} M. Jaminon, C. Mahaux and H. Ng\^{o}, Phys. Lett. {\bf 158B} (1985)
103
\bibitem{10} M. Jaminon, C. Mahaux and H. Ng\^{o}, Nucl. Phys. {\bf A473}
(1987) 509
\bibitem{11} A.N. Antonov, P.E. Hodgson and I.Zh. Petkov, {\it Nucleon Momentum
and Density Distributions in Nuclei}, Clarendon Press, Oxford, 1988
\bibitem{12} A.N. Antonov, P.E. Hodgson and I.Zh. Petkov, {\it Nucleon
Correlations in Nuclei}, Springer-Verlag, Berlin-Heidelberg-New York, 1993
\bibitem{13} P.-O. L\"{o}wdin, Phys. Rev. {\bf 97} (1955) 1474
\bibitem{14} O. Bohigas and S. Stringari, Phys. Lett. {\bf 95B} (1980) 9
\bibitem{15} H. Clement, P. Grabmayr, H. R\"{o}hm and G.J. Wagner, Phys. Lett.
{\bf 183B} (1987) 127
\bibitem{16} E.N.M. Quint, B.M. Barnett, A.M. van den Berg, J.F.J. van den
Brand, H. Clement, R. Ent, B. Frois, D. Goutte, P. Grabmayr, J.W.A. den Herder,
E. Jans, G.J. Kramer, J.B.J.M. Lanen, L. Lapik\'{a}s, H. Nann, G. van der
Steenhoven, G.J. Wagner and P.K.A. de Witt Huberts, Phys. Rev. Lett. {\bf 58}
(1987) 1088
\bibitem{17} D.S. Lewart, V.R. Pandharipande and S.C. Pieper, Phys. Rev. {\bf
B37} (1988) 4950
\bibitem{18} M.V. Stoitsov, A.N. Antonov and S.S. Dimitrova, Phys. Rev. {\bf
C47} (1993) R455
\bibitem{19} M.V. Stoitsov, A.N. Antonov and S.S. Dimitrova, Phys. Rev. {\bf
C48} (1993) 74
\bibitem{20} C. Ciofi degli Atti, E. Pace and G. Salm\`{e}, Phys. Rev. {\bf
C43} (1991) 1155
\bibitem{21} F. Malaguti, A. Uguzzoni, E. Verondini and P.E. Hodgson, Riv.
Nuovo Cim. {\bf 5} No.1 (1982) 1
\bibitem{22} A.N. Antonov, D.N. Kadrev and P.E. Hodgson, Phys. Rev. {\bf C50}
(1994) 164
\bibitem{23} K. Ypsilantis and M. Grypeos, Nuovo Cim. {\bf 82 A} (1984) 93
\bibitem{24} M.E. Grypeos and K. Ypsilantis, J.Phys. G {\bf 15} (1989) 1397
\bibitem{25} A.N. Antonov, M.V. Stoitsov, L.P. Marinova, M.E. Grypeos, G.A.
Lalazissis and K.N. Ypsilantis, Phys. Rev. {\bf C50} (1994) 1936
\bibitem{26} R.F. Frosch, J.S. McCarthy, R.E. Rand and M.R. Yearian, Phys. Rev.
{\bf 160} (1967) 874
\bibitem{27} I. Sick and J.S. McCarthy, Nucl. Phys. {\bf A150} (1970) 631
\bibitem{28} J.B. Bellicard, P. Bounin, R.F. Frosch, R. Hofstadter, J.S.
McCarthy, F.J. Uhrhane, M.R. Yearian, B.C. Clark, R. Herman and D.G. Ravenhall,
Phys. Rev. Lett. {\bf 19} (1967) 527
\bibitem{29} I. Sick, Phys. Lett. {\bf 53B} (1974) 15
\bibitem{30} M.V. Stoitsov, A.N. Antonov and S.S. Dimitrova, Z. Phys. {\bf
A345} (1993) 359
\bibitem{31} J.G. Zabolitzky, W. Ey, Phys. Lett. {\bf 76B} (1978) 527
\bibitem{32} J. Kramer, Ph.D. thesis, Amsterdam (1990) (unpublished)
\bibitem{33} H. de Vries, C.W. de Jager and C. de Vries, At. Data Nucl. Data
Tables {\bf 36} (1987) 495
\end{thebibliography}
\end{document}